\documentstyle[12pt]{article}
\def\pictleft#1#2#3#4#5{%
\begin{figure}[htbp]
\begin{minipage}[t]{#1}
\framebox[#1][l]{\raisebox{0cm}[0cm][#2]{ }}
\end{minipage}
\
\hfil
\
\raisebox{0.5cm}{\begin{minipage}[t]{#3}
\caption[]{\label{#4}\sloppy#5}
\end{minipage}}
\end{figure}
}
\def\pictup#1#2#3#4{%
\begin{figure}
\begin{center}
\framebox[#1][l]{\raisebox{0cm}[0cm][#2]{ }}
\end{center}
\caption[]{\label{#3}\sloppy#4}
\end{figure}
}
\def\pictuph#1#2#3#4{%
\begin{figure}[htbp]
\begin{center}
\framebox[#1][l]{\raisebox{0cm}[0cm][#2]{ }}
\end{center}
\caption[]{\label{#3}\sloppy#4}
\end{figure}
}
\begin{document}
\noindent
\hfill TTP93--18\\
\mbox{}
\hfill  June  1993   \\   
\vspace{0.5cm}
\begin{center}
  \begin{Large}
  \begin{bf}
  \vspace{3cm}
TOPICS IN TOP QUARK PHYSICS
 \footnote{\normalsize Supported by BMFT Contract 056KA93P}
\\
  \end{bf}
  \end{Large}

  \vspace{0.8cm}
  \begin{large}
   J.H. K\"uhn \\[5mm]
  \end{large}
    Institut f\"ur Theoretische Teilchenphysik\\
    Universit\"at Karlsruhe\\
    Kaiserstr. 12,    Postfach 6980\\[2mm]
    7500 Karlsruhe 1, Germany\\
  \vspace{4.5cm}
  {\bf Abstract}
\end{center}
\noindent
The status of top quark searches will be briefly reviewed.
Theoretical predictions for the top quark decay rate are
presented including QCD and electroweak radiative corrections.
The possibilities for quarkonium searches at an hadron collider
will be discussed.  The perspectives for top production at an
electron positron collider will be described in detail with
emphasis on the behavior of the cross section and decay
distribution in the threshold region.

\newpage
\section{Introduction}
The ever-increasing energies of electron
positron colliders from PETRA, PEP, and TRISTAN up to LEP as well
as those of the hadron colliders $Sp\bar pS$ and TEVATRON have led to
an impressive increase of the lower bound of the top quark mass.
The limits from LEP (44.5 GeV) and from the absence of
$W\to t+b$
as inferred from
the $W$ decay rate (54 GeV) are derived in a fairly
model-independent way.  The limits from the Tevatron on the other
hand depend on the dominance of the $W\to t+b$
decay mode and hence are
applicable in the context of the Minimal Standard Model (MSM)
only.  The mass bounds of 108 GeV from CDF and 103 GeV from D0
which have been quoted recently \cite{li} are derived on the basis of
production cross sections which in both cases are based on
next-to-leading-order calculations \cite{nlo}.  However, the D0 limit
includes recent higher-order contributions \cite{giele}
which raise the
cross section and increase the limit by about 5 GeV.

Precision measurements of electroweak parameters exploit the
quadratic top mass dependence \cite{veltman}
of the $\rho$-parameter, resulting  in \cite{schailering}
$m_t=155+17/-19{\rm GeV}$
with an additional error of $+17/-20$GeV
from the Higgs mass range between 60
GeV and 1 TeV with a central value at 200 GeV.  Future top
searches at the TEVATRON with increased statistics and efficiency
will able be to cover this interval to a large extent and should
finally be able to close the window.

Top masses above 54 GeV and below the Tevatron limit are still
possible in extensions of the MSM consistent with the present
data, notably in the two Higgs doublett model (THDM) where the
decay mode $t\to H+b$ may dominate for $m_t$ below  90 GeV.
This window will be closed by LEP200.

The upper limit from radiative corrections may be evaded in
models with negative contributions to the $\rho$-parameter such as
THDM with large mass splittings of the type $m_S\ll m_\pm\ll m_{ps}$.
This option
will, however, be eliminated with precise experimental results
for $\Gamma(Z\to b\bar b)$.

The discovery of the top quark will not only complete the fermionic
spectrum of the Standard Model.  The determination of its mass
will provide important input for all calculations, leaving the
Higgs mass as only free parameter which could then be inferred
from precise measurements planned at LEP in the near future
Fig.\ref{schaile}.  Once discovered, the study of the top quark and
its decay modes will constitute an important part of the physics
program at the TEVATRON, LHC, SSC, and the future linear collider.

Open top quark production will be covered in the talks by
Protopopescu, Harral and Froidedevaux.  I will therefore concentrate
on the perspectives for toponium production at the hadron
collider and on the possibilities at  a electron positron
collider \cite{rev}.

\section{Top decays}
Top quark physics is to a large extent dictated
by the large decay rate (Fig.\ref{schaile}),
which  quickly reaches its asymptotic form
\begin{equation}
\Gamma \left(t \to b+W^+\right) \to 175{\rm MeV}
\cdot \left(\frac{m_t}{
m_W}\right)^3
\label{eq:3}
\end{equation}
Three steps in the increase of the decay rate affect the
qualitative behavior of top mesons \cite{bigi}.\\
\pictleft{8cm}{8.5cm}{8.5cm}{schaile}%
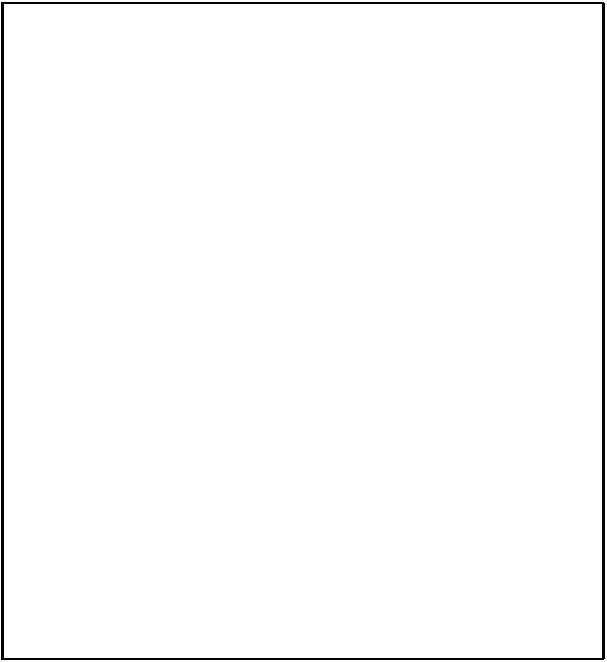
{Solid line: The width of the top quark in the SM, including
QCD and electroweak corrections and the finite $W$ width.
Dashed line: Born and narrow width approximation.}
\noindent
{\em i)}  Once $m_t$ is above 70 GeV the hyperfine splitting between
$T$ and $T^*$ cannot be resolved anymore:
\begin{displaymath}
m_{T^*} - m_{T} \propto 1/m_t < 1 {\rm MeV} < \Gamma_t
\end{displaymath}
The angular distribution of all decay products follow the
predictions derived for a spin 1/2 $t$-quark and may serve to
analyze the top spin.

\noindent
{\em ii)}  Above 100 GeV the width exceeds 100 MeV, the characteristic
scale for the formation of top mesons.  No $T$ or $T^*$ mesons can be
formed.

\noindent
{\em iii)}  Toponium states are severely affected once
$m_t$ is above 130 GeV, leading to a toponium width
$2\Gamma_t\ge0.88 {\rm GeV}$
comparable to the $1S$-$2S$ mass difference.
A complicated interplay between binding forces and the rapid decay
is predicted.
\begin{table}[h]
\begin{center}
\begin{tabular}{|r|c|c|c|c|c|c|c|c|c|} \hline
$m_t\ $  & $\alpha_s(m_{t})$ & $\Gamma^{Born}_{nw}$ &$\delta^{(0)}_{}$&
$\delta^{(1)}_{QCD}$
&$\delta^{}_{ew}$
&$\Gamma^{}_{t}$\\
{\scriptsize(GeV)} &  & {\scriptsize(GeV)} & {\scriptsize(\%)} &
{\scriptsize(\%)} &
{\scriptsize(\%)} &
{\scriptsize(GeV)} \\
\hline
 110.0& .115& .1955& -1.44 &-7.83&  1.20& .1796\\
 150.0& .110& .8852& -1.69 &-8.47&  1.57& .8087\\
 190.0& .106& 2.059& -1.39 &-8.47&  1.73& 1.890\\
\hline
\end{tabular}
\end{center}
\caption{Top width as a function of top mass and the comparison of
the different approximations.}
\end{table}

These quantitative considerations are supplemented by precise
calculations including QCD and electroweak corrections to the
decay rate and the spectra.  A reduction of $\Gamma_t$ by 7\% is
induced by QCD \cite{jezrate1,jezrate2}.
Electroweak corrections increase the rate by about
1.5\% \cite{weakrate}.
The finite width of the $W$ leads to reduction
by about the same magnitude \cite{jezrate2}.
The individual corrections and the
overall predictions are displayed in Table 1 for a few
characteristic masses.
QCD corrections including the effects from the nonvanishing $\Gamma_W$
and $m_b$ are available in the literature \cite{jezrate1,jezrate2}.
Several sources of theoretical uncertainties remain.  Not yet
calculated ${\cal O} (\alpha_s^2)$-terms
can be estimated to contribute about 1\%,
the influence of the uncertainty in $m_t$ (estimated to .5
GeV) will induce another 1\%.  With an estimated experimental
resolution of about 10\% at best at a future linear collider,
these uncertainties will not pose a serious problem in the
foreseeable future.

An important issue will be the search for new decay modes
expected in extensions of the Standard Model.  The impact of the
THDM on radiative corrections to $t\to W+b$ is fairly small
\cite{denhoang}. For a
top above 90 GeV with its large semiweak decay rate, new
additional modes will manifest as admixtures to the dominant
SM channel.  Branching ratios of about 10\% are expected for
$t\to H+b$ with
a plausible choice of the Yukawa coupling and similar or even
smaller numbers are predicted \cite{rev}
for $t\to\tilde t+\tilde g$. Exotic FCNC decays $t\to Z+c$ have
been proposed \cite{fcnc}
to exceed the SM prediction of $10^{-12}$ with the
branching ratios ranging up to $10^{-3}$ or even up to 1\%.

The lowest-order prediction for the energy and angular
distribution of leptons from a polarized top quark in the narrow
width approximation and with $m_b$ is extremely simple and
factorizes into an energy and an angular dependent term.  QCD
corrections to this formula are available in the literature
\cite{distcor}
and will be of use for future experimental studies.

\section{Toponium at hadron colliders}
The search for $\eta_t$ in its $\gamma\gamma$
decay mode has been recently proposed as a convenient tool for a
precise determination of $m_t$
at an hadron collider \cite{rub}.  The
small branching ratio around $10^{-4}$ or below, the tiny production
cross section and the large background are important obstacles.
The observable cross section in the $\gamma\gamma$ channel is
proportional
to the fourth power of the wave function at the origin and hence
depends strongly on the choice of the potential.  Realistic
calculations based on the two loop QCD potential and including
QCD corrections for production and decay have been performed
recently \cite{eta}.  They lead to results about a factor four below
the original estimates so that this reaction will be difficult to
observe in practice.  The chances are more promising for
quarkonium with suppressed single quark decays.  The decay of a
fourth generation $b'$  boundstate into $Z+H$
for example dominates all other channels,
and the resulting cross sections are large for a wide range of
masses \cite{exotic}.
This reaction could therefore lead to the simultaneous
discovery of $b'$ and the Higgs boson.
\section{Perspectives for $e^+e^-$ collisions}
{\em Top in the continuum}\\ \noindent
The cross section for $e^+e^-\to t\bar t$ is
fairly large throughout.  It starts with a step-function-like
increase to $R\approx0.7$
at threshold as a consequence of the Coulomb
enhancement $\propto1/v$
and increases to a value of $R\approx 2$ for energies
above $2m_t+100{\rm GeV}$ (Fig.\ref{rfig}).
\pictup{10.1cm}{4cm}{rfig}%
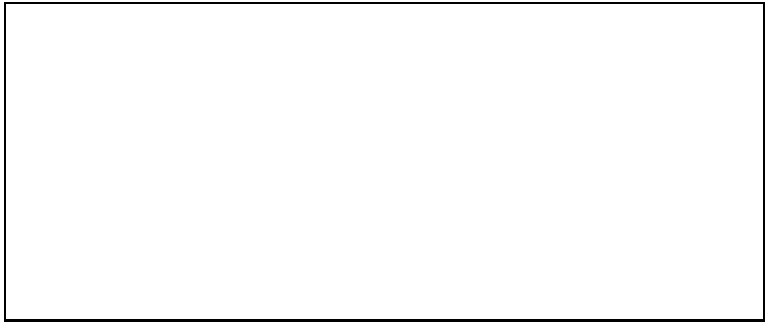
{Cross section for $t\bar{t}$ production,
including resonances, QCD corrections
and initial state radiation in units of $\sigma_{point}$.}

The production rate for $t\bar  t$
and the number of events (for $\int {\cal L} dt=10 {\rm fb}^{-1}$) are
contrasted with the potential backgrounds $f\bar f$, $WW$ and $ZZ$
in Table 2.
Detailed simulations demonstrate \cite{pik}
that a ratio between signal
and noise of 10 can be achieved with an efficiency of 30\%.  The
top mass can be determined with an estimated uncertainty of about
0.5 GeV.

The large top decay rate allows for a variety of novel
QCD studies. The rapid decay intercepts the evolution of the
string of soft hadrons between $t$ and $\bar t$ \cite{sjostr}.
For an extremely large top width of more than 2 GeV
corresponding to $m_t$ close to 200 GeV the string is actually
spanned between the $b$ and the $\bar b$
such that the multiplicity depends on the
invariant mass of the $b\bar b$ system.  Perturbative soft gluon
radiation is similarly affected and cut off for particular kinematic
configurations \cite{soft}.  This effect is particularly pronounced
for $e^+e^-$ energies in the TeV range.
\begin{table}[htbp]
\begin{center}
\begin{tabular}{||c||c|c|c|c|c||}
   \hline  \hline
Top mass (GeV)          & 120 & 150 & 180 & 150 & 150 \\
$E_{c.m.}$ ($GeV$)  & 250 & 310 & 370 & 400 & 500 \\
   \hline \hline
$t\bar t$    & 0.972 & 0.632 & 0.444 & 0.821 & 0.656 \\
$f \bar f$ & 58.6 & 36.6 & 25.6 & 21.9 & 16.6 \\
$W^+W^-$  & 15.2 & 12.5 & 10.3 & 9.46 & 7.44 \\
$Z^0Z^0$ & 1.25 & 0.983 & 0.792 & 0.719 & 0.572\\
   \hline \hline
\end{tabular}
\end{center}
\caption[]{\label{Tbackgr}
Cross sections in {\it picobarn}
for the $t\bar t$ signal and for the main
background processes. When $E_{cm}$ is close to the threshold (first
3 columns), the $t\bar t$  cross section is
taken as 0.7 times the ``point cross section''. Far above threshold,
the cross sections are those calculated by PYTHIA. All cross sections
are obtained with the inclusion of
initial state radiation.}
\end{table}

\vspace{3mm}
\noindent
{\em Cross section in the threshold region}  \\ \noindent
A lot of
effort has been invested recently in the study of the total cross
section and of momentum distributions in the threshold region.
For $m_t$ above 120 GeV
one faces a series of many overlapping resonances.
A convenient technique to perform the summation has been proposed
in \cite{khoze}.
The sum over the individual Breit-Wigner resonances
multiplied by the wave function at the origin
is easily transformed into the imaginary part of the Green's
function.

\begin{equation}
       \sum_{n} \left| \psi_n (0) \right|^2
       \frac{\Gamma_t}{(E_n - E)^2 + \Gamma^2_t} =
       \sum_{n} \mbox{Im} \frac{\psi^{\*}_n (0) \psi^*_n (0)}
       {E_n - E - i \Gamma_t}=
       \mbox{Im} G(0,0,E+i\Gamma)
\end{equation}

The latter can be calculated analytically for the Coulomb
and numerically
for arbitrary potentials \cite{peskin}.

The total cross section is affected by initial state radiation
that is accounted for by the standard radiator function.
Electroweak corrections affect the cross section by an
overall factor that can be calculated independently of the
potential and other QCD effects as a consequence of the short
distance nature of $W$ and $Z$ exchange.
Only the exchange of a light
Higgs boson with large couplings may exhibit a more complicated
interplay with the potential.  The leading correction is in this
case written as a factor
\begin{displaymath}
1+\frac{\sqrt{2} G_F m_t^2}{4\pi}\frac{m_t}{m_H}
\end{displaymath}
and can be traced to Coulomb-like singularity from an
instantaneous Yukawa potential.
\begin{displaymath}
V(r)=\frac{\sqrt{2} G_F m_t^2}{4\pi}\frac{e^{-r m_H}}{r}
\end{displaymath}

For realistic masses $m_t\approx150{\rm GeV}$ and $m_h> 60 GeV$
retardation effects must be
incorporated which are in this
case as important as the $1/m_t$ term \cite{guth}.
A combined treatment incorporating both the instantaneous
potential and the retardation effects is given in \cite{tobe}.
\pictuph{10.5cm}{6.7cm}{mom}{%
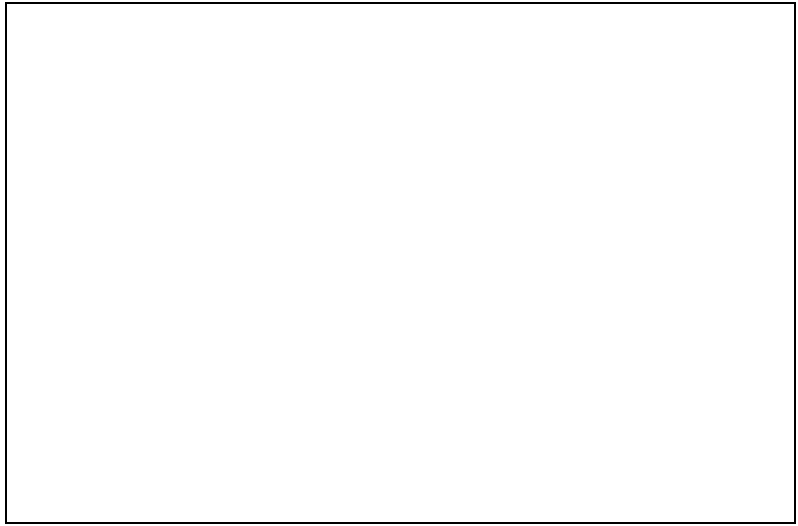
Green's function $|p{\cal G}(p,E+i\Gamma_t)|^2$ for
$m_t = 120$  and  $E= -2.3$  GeV
(1s peak) -- solid, $E = 0$ --
dotted, and $E = 2$ GeV -- dashed line.}
\\
\vspace{3mm}
\noindent
{\em Momentum distributions}
\\ \noindent
The momentum distribution of
top quarks from an individual resonance can be calculated from
the Fourier transform of the wave functions.
\begin{equation}
\frac{dN}{d\vec p} \bigg|_{E=E_n} = |\tilde\psi_n(\vec p)|^2
\end{equation}
For a series of overlapping resonances, the momentum distribution
of top quarks and the decay products is again conveniently
evaluated with the help of Green's functions:
\begin{equation}
{d\sigma\over d\vec{p}} \left(\vec p,E\right) =
{3\alpha^2Q_t^2\over \pi sm_t^2}
\Gamma_t\left|{\cal G}(\vec{p},E+i\Gamma_t)\right|^2
\label{eq:md3}
\end{equation}
$G(\vec p,E=i\Gamma)$ can be calculated
numerically for complex energies and an
arbitrary QCD potential by solving the Lipmann-Schwinger equation
\begin{equation}
{\cal G}(\vec{p},E+i\Gamma_t) =
{\cal G}_0(\vec{p},E+i\Gamma_t) +
{\cal G}_0(\vec{p},E+i\Gamma_t)
\int {d\vec{q}\over(2\pi)^3}
\tilde V(\vec{p}-\vec{q})
{\cal G}(\vec{q},E+i\Gamma_t)
\label{eq:md5}
\end{equation}
where $\tilde V(\vec{p})$ is the potential in momentum space.
The free Hamiltonian that is used to define the Green's function
${\cal G}_0$ includes the width.
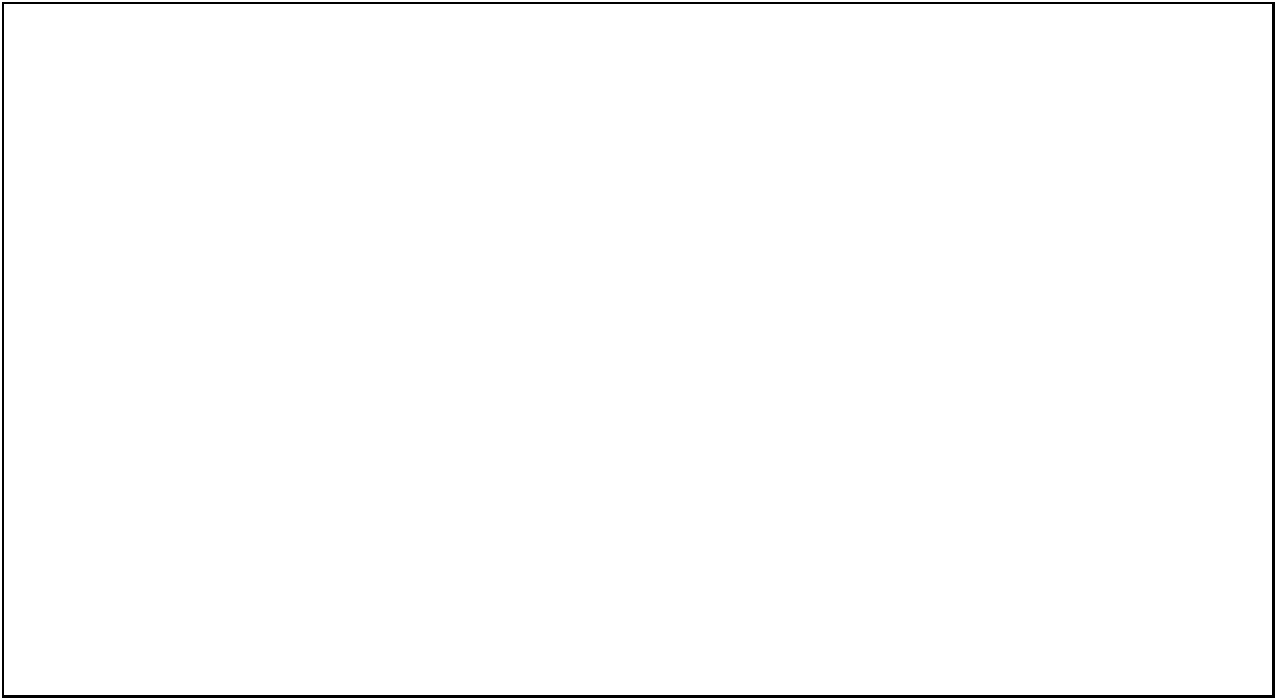
\pictuph{16.9cm}{9cm}{momav}{{\em Left:}
Energy dependence of $\langle p\rangle$, the
average t quark momentum
for $\alpha_s = 0.13$ (dotted)
0.12 (dashed) and 0.11 (solid) line for
$m_t=120$ GeV  and
$\Gamma_t=0.3 {\rm GeV}$.\\
{\em Right:} Average momentum for $\alpha_s=0.125$ and
different top masses.}
The resulting momentum distribution is shown in Fig.\ref{mom} for a
characteristic set of energies.  The average momentum as a
function of energy (Fig.\ref{momav}) depends on $\alpha_s$ for
negative energies, that is in
the binding region.  Above threshold the momentum simply reflects
the kinematic behavior $p=\sqrt{m_t E}$
and can be used to measure $m_t$
independently from $\alpha_s$.
Up to this point final state interactions
and the reduction of phase space by the $t\bar t$ binding have been
neglected.  At first sight one might expect a drastic reduction
of the decay rate by about 10\% from the reduction of the phase
space through
the potential energy of about 5 GeV.  However, as argued
in \cite{jezt} this is largely compensated by an enhancement of the
rate through final state interaction as a result of the Coulomb wave
function of the outgoing $b$ quark.  A slight reduction of the
rate through time dilation
leads to a completely negligible change in the total cross
section and the momentum distribution.  Even the most pessimistic
assumptions about these ${\cal O}(\alpha_s^2)$
corrections will, however, leave the
shape of the threshold cross section unaffected, leading to a
stable determination of the top mass.

{\em In summary:}  An $e^+e^-$
collider will lead to precise determination of
the top mass.  Top studies will provide an exciting laboratory
for QCD.  The determination of the $Ht\bar t$ Yukawa coupling will be of
prime importance.  The $Wtb$ gauge coupling can be measured and
limits on anomalous couplings can be derived.  Novel decay modes
could be searched and studied in a clean environment.

\end{document}